# Photophysics of Two-Dimensional Semiconducting Organic-Inorganic Metal-Halide Perovskites

Daniel B. Straus[1] and Cherie R. Kagan[2]*

[1]Department of Chemistry, Princeton University, Princeton, NJ 08544 USA

[2]Departments of Electrical and Systems Engineering, Materials Science and Engineering, and Chemistry, University of Pennsylvania, Philadelphia, PA 19104 USA

*Author to whom correspondence should be addressed. Email: kagan@seas.upenn.edu

**Abstract**

2D organic-inorganic hybrid perovskites (2DHPs) consist of alternating anionic metal-halide and cationic organic layers. They have widely tunable structural and optical properties. We review the role of the organic cation in defining the structural and optical properties of 2DHPs through example lead iodide 2DHPs. Even though excitons reside in the metal halide layers, the organic and inorganic frameworks cannot be separated—they must be considered as a single unit to fully understand the photophysics of 2DHPs. We correlate cation-induced distortion and disorder in the inorganic lattice with the resulting optical properties. We also discuss the role of the cation in creating and altering the discrete excitonic structure that appears at cryogenic temperatures in some 2DHPs, including the cation-dependent presence of hot exciton photoluminescence. We conclude our review with an outlook for 2DHPs, highlighting existing gaps in fundamental knowledge as well as potential future applications.

# 1. Introduction

Semiconductors have an electronic structure that is "just right," with band gaps tailorable through the optical spectrum by composition and dimension and carrier concentrations that can be modulated by external stimuli.(1) The properties of semiconductors are of fundamental scientific interest—to understand matter and its interactions in particular with light and voltage.(2) They are also harnessed in electrical and optoelectronic devices, for applications in sensing, energy, computation, and communications technologies.(3) Conventional semiconductors are extended inorganic materials, namely elemental group IV; compound III-V, II-VI, and IV-VI; and ternary and quaternary compositions, and they may be bulk solids or low-dimensional homo- or hetero-structures. Organic materials such as aromatic molecular crystals as well as oligomeric and polymeric solids may also be semiconducting.(4) Finally, organic-inorganic hybrid materials also form semiconductors. They combine the properties of organic and inorganic materials, greatly increasing their complexity both in the number of possible compositions and in their physical properties. These materials include assemblies of colloidal nanostructures(5, 6) as well as metal-halide hybrid perovskites,(7, 8) which are the focus of this review.

Inorganic metal-halide perovskites are stoichiometric compounds with the chemical formula $AMX_3$ (Figure 1A).(9–12) A singly-charged cation ($A^+$; green in Figure 1A) occupies the interstitial space in a corner-sharing network of $MX_6$ octahedra, with M being a 2+ metal (typically $Pb^{2+}$ or $Sn^{2+}$; grey in Figure 1A) and X a 1- halide ($Cl^-$, $Br^-$, or $I^-$; purple in Figure 1A). The size of the A-site cation is limited by the available volume of the interstitial space between the $MX_6$ octahedra—if $A^+$ is too small or too large, a 3D perovskite cannot form. This principle is formalized through the Goldschmidt tolerance factor,(13, 14) which is derived from the principle that an ionic structure is stable when the number of cation-anion contacts is maximized and free-volume within a structure is minimized.(15) Based on the available volume in the voids between the $MX_6$ octahedra, the only inorganic cation that forms a stable perovskite structure with $Pb^{2+}$ or $Sn^{2+}$ is $Cs^+$, the largest 1+ inorganic cation.(16) Even $Cs^+$ is too small for $CsPbI_3$ to be a stable perovskite.(17)

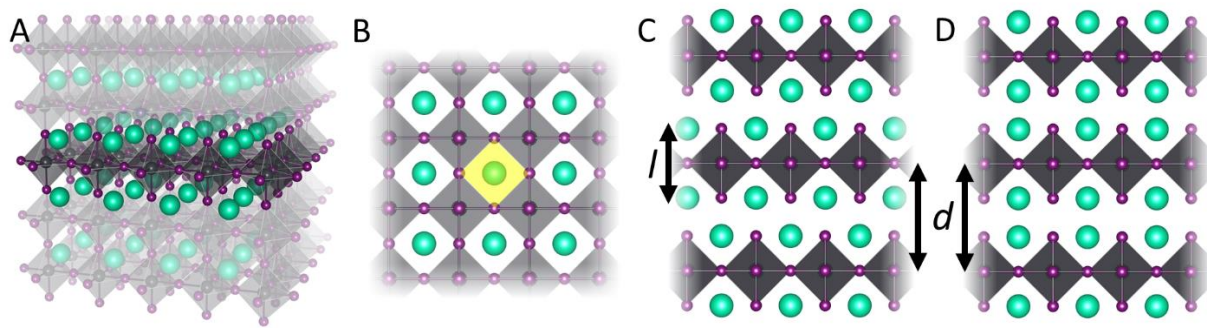

**Figure 1: Constructing 2DHPs.** A) Cubic 3D halide perovskite with the formula $AMX_3$, with $A^+$ as green, $M^{2+}$ as grey, and $X^-$ as purple spheres. $MX_6$ octahedra are shaded in grey. 2DHPs are constructed by taking a single $A_2MX_4$ (darker atoms) layer from the overall 3D structure. B) Top view of a single layer of a 2DHP, with the maximum cross-sectional area highlighted in yellow. Different stacking arrangements are possible in 2DHPs. C) The Ruddlesden-Popper phase, where each layer is offset from one another by half a lattice constant and D) the Dion-Jacobson phase, where there is no offset between layers in at least one direction. The interlayer spacing $d$ is the distance between planes of Pb atoms in adjacent inorganic layers (black arrows), and the thickness of the inorganic layer $l$ is the distance between axial X- atoms in an octahedron.

Organic-inorganic hybrid perovskites replace the $A^+$ inorganic cation with a positively charged organic molecule, greatly increasing the phase-space of possible materials.(7) In a three-dimensional (3D) hybrid perovskite (3DHP), the organic cation must be small, and the most popular choice is methylammonium ($CH_3NH_3^+$). First discovered in 1978,(18, 19) hybrid perovskites attracted tremendous attention starting in the mid-2000s when they were first used for solar energy conversion.(20, 21) The efficiency of hybrid perovskite solar cells has rapidly increased, now rivaling that of commercial silicon-based solar-cells.(22)

While increasing the total size of the cation prevents the perovskite structure from forming, increasing the length of the cation results in the spontaneous formation of two-dimensional (2D) hybrid perovskites (2DHPs).(23) A typical 2DHP can be constructed by cutting a single layer of corner-sharing octahedra out of a 3D perovskite, including the cations on both sides (Figure 1A), and then stacking these layers on top of each other. This results in a material with the formula $A_2MX_4$. A top-down view of a single layer of a 2DHP is shown in Figure 1B. $A_2MX_4$ 2DHPs are van der Waals materials, and the stacking of 2DHP layers in thin films or crystal(lites) can occur in multiple ways.(24–26) If the individual layers are offset from one another by half a lattice constant in both directions, they are referred to as Ruddlesden-Popper phase (Figure 1C). Conversely, if there is no offset in one or both directions, the Dion-Jacobson phase occurs (Figure 1D). Both Ruddlesden-Popper and Dion-Jacobson 2DHPs have been synthesized.(25) While a cation with too large a cross-sectional area (larger than yellow region in Figure 1B) will disrupt 2DHP formation, there are no known restrictions on the length of the cation.(7) The flexibility in the choice of cation provides tremendous structural diversity in 2DHPs,(24, 25) with the ability to tune the structure by the choice of cation within the limit of the cross-sectional area of the cation, depicted in yellow in Figure 1B.

This review limits its consideration to monolayer 2DHPs, where the inorganic framework is a single layer of the parent 3DHP material (Figure 1A) and the shared corners of the octahedra are all located in a single plane (Figures 1B-C). We do not discuss examples with alternate octahedral connectivity or thicker inorganic frameworks (i.e., with multiple, 2 or more, inorganic layers per organic cationic layer).(7) We also limit our consideration to 2DHPs where the states that originate from the A-site cations are far from the edges of the metal-halide valence and conduction bands. 2DHPs with this electronic structure confine carriers and excitons to the inorganic framework. This electronic structure is known as a Type I band alignment, which is also seen in semiconductor quantum well superlattices.(24, 27) The most studied, inorganic framework is composed of <100> Pb-I inorganic layers, which we will focus on in this review to exemplify the photophysics of 2DHPs.

Common to the study of all semiconductors is the importance of optical spectroscopy, a key tool of physical chemists, in studying the electronic and vibrational structure of and the dynamics of excitons, electrons, and phonons in semiconductors.(2, 4, 28) *In this review, we highlight how the optical properties of excitons in 2DHPs can be tuned through the choice of cation without varying the metal or halide.* We review several situations where the organic cation is required to understand and model the properties of excitons in 2DHPs and explore how changes in the composition, length, and cross-sectional area of the cation influence the photophysical properties of excitons. Section 2 discusses the electronic structure of 2DHPs, examining the impact of quantum and dielectric confinement effects on the 2DHP semiconductor band structure and on the energy of excitons. Section 3 illustrates the impact of cation-induced structural changes on the optical and electronic properties of 2DHPs. In Section 4, we examine how longer cations impact the energy landscape of 2DHPs. Section 5 considers the origin of the low-temperature structure in the excitonic absorption and photoluminescence resonances, focusing on the hypothesis that the organic cation is responsible for this structure. Lastly, Section 6 gives an outlook on future directions in the study of 2DHPs.

**2. Modeling the properties of 2DHPs**

**2.1 Band structure**

The band structure in solid-state theory describes the dispersion relationship between the energy and wavevector ($k$) of an electron.(29) Conventional semiconductors have a valence band derived from atomic

*p*-orbitals and a conduction band derived from atomic *s*-orbitals yielding more complex, near-degenerate heavy and light hole as well as spin-orbit split-off valence bands but only a single conduction band.(3, 30) In contrast, metal-halide perovskites have a band structure different from that of conventional semiconductors. The valence band of halide perovskites is composed of antibonding orbitals formed from halide *p*-orbitals and metal *s*-orbitals, while the conduction band is predominantly derived from antibonding orbitals composed of metal *p*-orbitals (Figure 2A).(31, 32) This band structure leads to one valence band but complexity in the conduction band, including significant spin-orbit coupling effects, which are especially significant for structures containing heavy atoms such as Pb and Sn (Figure 2B).(33) This is true in both 3D and 2DHPs for the most commonly studied aliphatic or short aromatic (e.g., phenyl or naphthyl) ammonium A-site cations,(34) where states originating with the A-site cation are far from the metal-halide band edges. The 2DHPs therefore have a Type I band alignment. Thus, in 2DHPs it is common to consider the inorganic and organic sublattices separately.(31, 34) Under this simple model, the composition of the metal and halide of the inorganic sublattice defines the conduction and valence band energies and the organic sublattice acts solely as an insulating spacer between the inorganic metal-halide layers. At this level of theory, 2DHPs are functionally equivalent to semiconductor quantum well superlattices prepared by molecular beam epitaxy, with the advantage that 2DHPs are free from interfacial roughness.(24, 27)

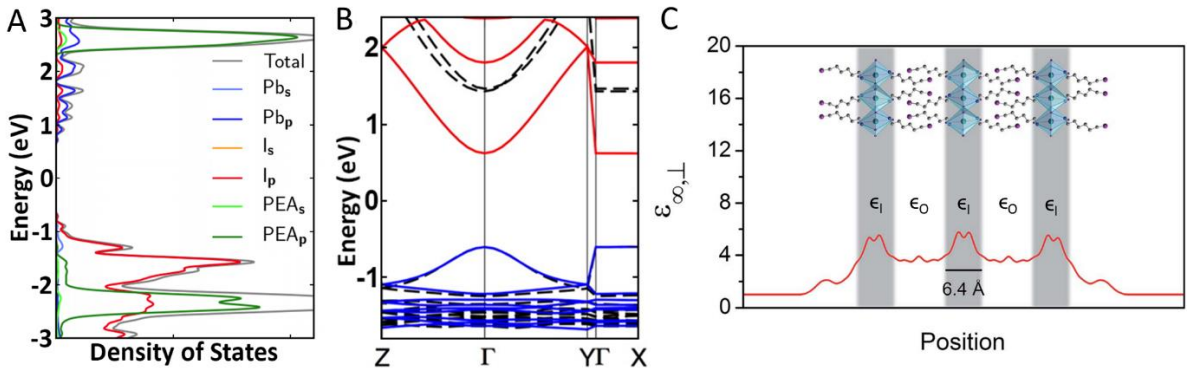

**Figure 2: Electronic structure.** A) Density of states for $(PEA)_2PbI_4$. Adapted with permission from ref (32). Copyright 2017 American Chemical Society. B) Band structure for $(4\text{-}FPEA)_2PbI_4$ without (dashed black) and with (blue and red) spin-orbit coupling. Adapted with permission from ref (35). Copyright 2016 American Chemical Society. C) Dielectric profile perpendicular to Pb-I layers of $(IC_6H_{12}NH_3)_2PbI_4$. Reproduced from ref (36) with permission from the Royal Society of Chemistry.

In 2DHPs, the inorganic framework extends infinitely in two directions, but the thickness *l* of the inorganic framework (Figure 1C) is restricted. For example, the thickness of a layer in lead iodide 2DHPs is only ~6.3 Å.(37) While the band structures of 3D and 2DHPs have the same atomic orbital contributions and are thus similar, in 2DHPs strong quantum confinement effects are present in one dimension, which increases their band gap compared to that of similar 3DHPs.(38) Assuming the organic framework imposes an infinite confinement potential, the valence (conduction) band of the 2DHP decreases (increases) by $\frac{\hbar^2\pi^2}{2m_{e,h}l^2}$, where $l$ is the thickness of the inorganic framework and $m_{e,h}$ is the mass of the hole (electron).(31)

The layered structure also gives rise to dielectric contrast between the high dielectric constant inorganic framework $\epsilon_I$ (grey, Figure 2C) and the lower dielectric contrast layer of organic cations $\epsilon_O$ (white, Figure 2C). This contrast gives rise to dielectric confinement effects. Because the inorganic layers are thin compared to the Bohr exciton radius,(39) the electric field generated by a charge carrier located in the inorganic framework extends into the organic framework, where the screening of the field is weaker.(40, 41) The reduction in dielectric screening increases the Coulomb potential in the inorganic layers increasing the band gap.(42)

## 2.2 Excitons

In 3D perovskites the excitonic resonance is not well-separated from the band-edge absorption at room temperature. 3D lead iodide perovskites such as methylammonium lead iodide ($CH_3NH_3PbI_3$) and cesium lead iodide ($CsPbI_3$) have binding energies of 12-16 meV and 18 meV respectively,(43, 44) so at room temperature ($kT$ = 26 meV), all excitons spontaneously separate into free electrons and holes. In contrast, Type I 2DHPs have a sharp excitonic absorption resonance that is well-separated from the onset of the continuum absorption (Figure 3A).(38) Minimal electronic coupling between adjacent inorganic layers is found in most 2DHPs.(45) Quantum and dielectric confinement greatly increase the exciton binding energy in 2DHPs by over an order of magnitude compared to their parent 3D materials, resulting in exciton binding energies that typically exceed 150 meV.

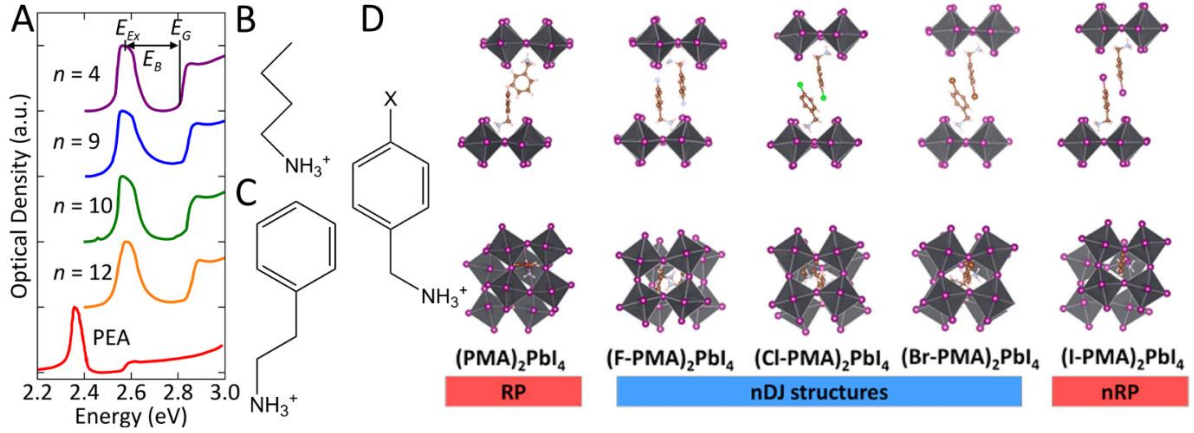

**Figure 3: Quantum and dielectric confinement effects.** A) Absorption spectra of $(C_nH_{2n+1}NH_3)_2PbI_4$ taken at 1.6 K and of $(PEA)_2PbI_4$ taken at 10 K, with the exciton energy $E_{Ex}$, band gap $E_G$, and exciton binding energy $E_B$ labeled. Data from refs (46, 47). B) Structure of $C_4H_9NH_3^+$, a representative alkylammonium cation. C) Structure of PEA cation. D) Derivatives of the phenylmethylammonium ($C_6H_5CH_2NH_3^+$, abbreviated PMA) cation can be used to tune the stacking motif between 2DHP layers from Ruddlesden-Popper (RP) to Dion-Jacobson (DJ). Reprinted with permission from ref (26). Copyright 2019 American Chemical Society.

In a perfect quantum well that is infinitely thin and has an infinite confinement potential, quantum confinement effects increase the binding energy such that $E_b = 4E_0$ where $E_0 = \left(\frac{1}{\epsilon_1}\right)^2 \left(\frac{m_{ex}}{m_0}\right) R_H$ is the exciton binding energy of the parent bulk 3D material, $m_{ex} = (1/m_e + 1/m_h)^{-1}$ is the reduced mass of the exciton, $m_0$ the mass of an electron, and $R_H$ the hydrogen Rydberg constant.(48) Because the inorganic layers in 2DHPs are not infinitely thin and the confinement potential is finite, this serves as an upper bound on the true quantum confinement-induced exciton binding energy enhancement. Quantum confinement would account for ~60 meV of the 2DHP exciton binding energy based on the ~15 meV binding energy in lead iodide 3DHPs.(43, 44) Thus, quantum confinement alone cannot explain the enhancement in exciton binding energy.

Dielectric confinement effects account for the rest of the enhancement in 2DHP's exciton binding energy compared to 3DHPs. The Coulomb potential that binds the electron and hole together as an exciton is greatly increased by dielectric confinement because the effective dielectric constant is smaller than that of the inorganic framework. The approximate increase in the exciton binding energy established by dielectric confinement is

$$\Delta E_b = 2 \frac{\epsilon_I - \epsilon_O}{\epsilon_I + \epsilon_O} \frac{e^2}{\epsilon_I l} A \qquad \text{(Eq. 1)}$$

where *e* is the fundamental charge, *l* is the thickness of the inorganic layer, and *A* is a factor that depends on *l* and is slightly less than unity; the full equation is given in ref (48).

**2.3 Nonidealities in real 2DHPs**

In conventional inorganic quantum well superlattices, the one-dimensional periodic potential established by the semiconductor layers is captured by the Kronig-Penney model using the constituent semiconductor effective masses and requiring continuity of the wave-functions at their interfaces.(49) This simple model is successful for example superlattices where the constituent semiconductors are chemically similar, the interfaces are coherent, and the Bloch functions (that capture the periodic potential of the unit cell) of the bulk material are at high symmetry points, so that the potential variation can be described by slow varying envelope functions and the atomic-scale potentials can be ignored. However, this simplified approximation breaks down when the semiconductors in a superlattice have very different electronic properties and their interfaces become structurally and electronically complex, such as having differences in local symmetry and electronic mixing between states at different points of the Brillouin zone.(31)

In 2DHPs, the organic and inorganic layers are chemically very different. While treating the organic and inorganic frameworks separately allows for an intuitive understanding of how changes to the composition and structure of 2DHPs affect their band structure and exciton behavior, this simple model is not surprisingly inaccurate and atomic-scale detail is necessary to capture the effects of quantum confinement. Even *et al.* showed that the electronic properties of 3DHPs cannot be used to model the inorganic framework in 2DHPs.(31) Charge carriers in 2DHPs are located at k-points away from the zone center where the bands of bulk 3DHPs are not parabolic, and thus the 3DHP effective masses (found from the band curvatures near the extrema) are not valid.(31, 38, 50) Assuming the organic framework imposes an infinite confinement potential is also not appropriate and overestimates the confinement potential. However, even imposing a finite confinement potential with a sharp transition between the organic and inorganic layers is also not sufficient to accurately model quantum confinement effects in 2DHPs.(31) The Coulomb interactions between the cationic organic and anionic inorganic layers and the atomic-scale detail of the interfaces are needed.

The quantum confinement potential has been quantitatively modeled by considering 2DHPs as composites that capture their ionic character.(31) For the representative 2DHP $(C_{10}H_{21}NH_3)_2PbI_4$, the inorganic sublattice is modeled as $Na_2PbI_4$, where the two $Na^+$ cations compensate for the negative charge of the $[PbI_4]^{2-}$ framework. The organic sublattice is modeled as $C_{10}H_{21}CH_3$, with the terminal N replaced by C to maintain charge neutrality of the combined $(C_{10}H_{21}CH_3)Na_2PbI_4$ model system. This method is shown to provide a good description of the electronic states close to the band edges, which are largely located in the inorganic layers, for the commonly explored 2DHPs with weak interactions between inorganic layers. It also describes the consistency in the electronic structure and energy gap of 2DHPs with similar inorganic structures.

While modeling the 2DHP superlattice in this more complex way accounts for quantum confinement, it is still too simplistic to capture the dielectric modulation and thus the effects of dielectric confinement on the electronic structure and exciton binding energy. Like the confinement potential, the dielectric constant is also more complex and is not a simple step function where it adopts a fixed value in the inorganic layer and another fixed value in the organic layer. Instead it is a smoothly-varying function (red, Figure 2C) and care must be taken to model it accurately, especially in the presence of intercalants (Section 2.5) or cations that result in an anisotropic dielectric profile where the parallel and perpendicular components of the dielectric function take on different values.(36)

Models that consider the organic and inorganic frameworks separately also do not capture the strong influence of the organic cation on the static and dynamic structure of the inorganic framework. The size and shape of the organic cations influence the structure of the inorganic sublattice, which in turn changes the band structure and affects how excitons and carriers behave (Section 3).(51) The influence of dynamic effects can be understood intuitively by considering the ionic nature of the material and the electrostatic interactions between the organic and inorganic sublattices—any motion of the positively charged ammonium group will alter the electrostatic potential in the inorganic framework. Even if we consider a high-energy vibrational mode of the 2DHP, in which only the cation moves because vibrations involving the heavy Pb and I atoms are low in energy,(52) the electric field created by the positively charged organic cation will be modulated by this motion. These and other subtleties will be addressed in Sections 3-5.

**2.5 Experimental manifestations of quantum and dielectric confinement**

The principles discussed in Sections 2.1-2.4 are immediately apparent in optical spectra of 2DHPs. The large binding energy of the exciton and thus large separation of its absorption resonances from the band edge is a signature of the strong contributions of quantum and dielectric confinement in 2DHPs. For commonly studied alkylammonium lead iodides $(C_nH_{2n+1}NH_3)_2PbI_4$ (Figure 3B), as n increases from 4 to 12, the interlayer Pb-Pb distances (*d* in Figure 1C-D) increase from 15.17 to 24.51 Å. The exciton binding energy $E_B$ is 320 ± 30 meV and the band gap $E_G$ is ~2.88 eV at 1.6 K for all of these 2DHPs (Figure 3A).(47) $(C_3H_7NH_3)_2PbI_4$ also forms a 2DHP with an interlayer Pb-Pb distance of 13.81 Å. Its optical properties do not differ from those of $(C_4H_9NH_3)_2PbI_4$, but unlike the other alkylammonium 2DHPs, $(C_3H_7NH_3)_2PbI_4$ converts to a non-perovskite phase upon exposure to moisture.(53) $C_2H_5NH_3^+$ does not form a 2DHP with Pb and I and instead forms 1D chains of face-sharing Pb-I octahedra,(54) and $CH_3NH_3PbI_3$ is the prototypical 3DHP. All $(C_nH_{2n+1}NH_3)_2PbI_4$ 2DHPs therefore experience a similar degree of quantum confinement, and a minimum critical length for the presence of full quantum confinement effects in alkylammonium 2DHPs cannot be experimentally established. It is important to note that because the optical properties do not change in these 2DHPs, the inorganic layers do not or at most weakly interact with one another (i.e., there are no superlattice effects).(31) 2DHPs with shorter cations have been synthesized,(55) and theoretical studies on the 2DHP $(IC_2H_4NH_3)_2PbI_4$ indicate it shows a reduced degree of quantum confinement compared to $(C_nH_{2n+1}NH_3)_2PbI_4$ 2DHPs. Unlike most 2DHPs, it is expected to exhibit weak superlattice effects.(31, 45)

Dielectric confinement effects can be used to tune the exciton binding energy even if the degree of quantum confinement is invariant. As the difference between $\epsilon_I$ and $\epsilon_O$ decreases, the exciton binding energy decreases according to Eq. 1. This can be seen in Figure 3A—the conjugated phenethylammonium $(C_6H_5C_2H_4NH_3^+$, abbreviated PEA; shown in Figure 3C) cation has a larger dielectric constant than unconjugated alkylammonium $(C_nH_{2n+1}NH_3^+)$ cations, and accordingly $(PEA)_2PbI_4$ has a smaller exciton binding energy of 190-220 meV (red, Figure 3A).(46, 52) It also has a smaller band gap, though some of the change in band gap can be attributed to the presence of additional strain in the inorganic framework (Section 3).

Dielectric confinement effects can be exploited to tune the exciton binding energy in a single 2DHP without synthetically modifying the cation. Intercalating molecular $I_2$ in the organic framework of the 2DHP $(IC_6H_{12}NH_3)_2PbI_4$ increases $\epsilon_O$. After the intercalation process, the band gap decreases by 70 meV from 2.56 to 2.49 eV, and the exciton binding energy decreases by 50 meV from 230 to 180 meV.(36) The molecular $I_2$ can be deintercalated, returning the band gap and exciton binding energies to the original values for $(C_6H_{12}NH_3)_2PbI_4$. In addition, when a single layer of the 2DHP butylammonium lead iodide $((C_4H_9NH_3)_2PbI_4)$ is exfoliated and placed on a silica ($\epsilon$ = 2.13) substrate with air ($\epsilon$ = 1) on top, the exciton binding energy $E_B$ increases to 490 meV for the monolayer from ~320-370 meV in the multilayer 2DHP.(56)

The stacking motif can be varied by tuning intermolecular interactions between the cations (Figure 3D)(26, 57) or by using more rigid bidentate cations,(58) where a single $A^{2+}$ cation has a positively charged ammonium functional group at each end, for an overall formula of $AMX_4$. Because these cations are all longer than $C_3H_7NH_3^+$, there is no difference in the optical or electronic properties of the 2DHPs shown in Figure 3D that can be attributed to differences in the stacking motif.(26) It is possible that subtle spectral changes may emerge at cryogenic temperatures from differences in stacking arrangement.(58)

### 3. Cation-induced changes to the inorganic framework

Thus far we have neglected to consider how changes to the inorganic framework's structure also affect the optical properties of 2DHPs. The model perovskites in Figure 1 exhibit no distortion—the $PbI_6$ octahedra have equal Pb-I bond lengths and 90° internal I-Pb-I bond angles. In addition, the octahedra are not tilted with respect to one another—the bridging Pb-I-Pb bond angles (θ in Figure 4A) are 180°.

Real 2DHPs display some degree of distortion (Figure 4B)—we are unaware of any experimental examples with completely undistorted cubic inorganic frameworks. The $PbI_6$ octahedra are tilted with respect to one another, and accordingly, θ is less than 180°. Even in distorted 2DHPs, the inorganic layers are usually planar (Figure 4C). However, in highly distorted 2DHPs, the inorganic framework buckles and the inorganic layers appear corrugated (Figure 4D). This section discusses how the degree of distortion can be tuned through cation modification, as well as the effect of inorganic lattice distortions on the optical properties of 2DHPs.

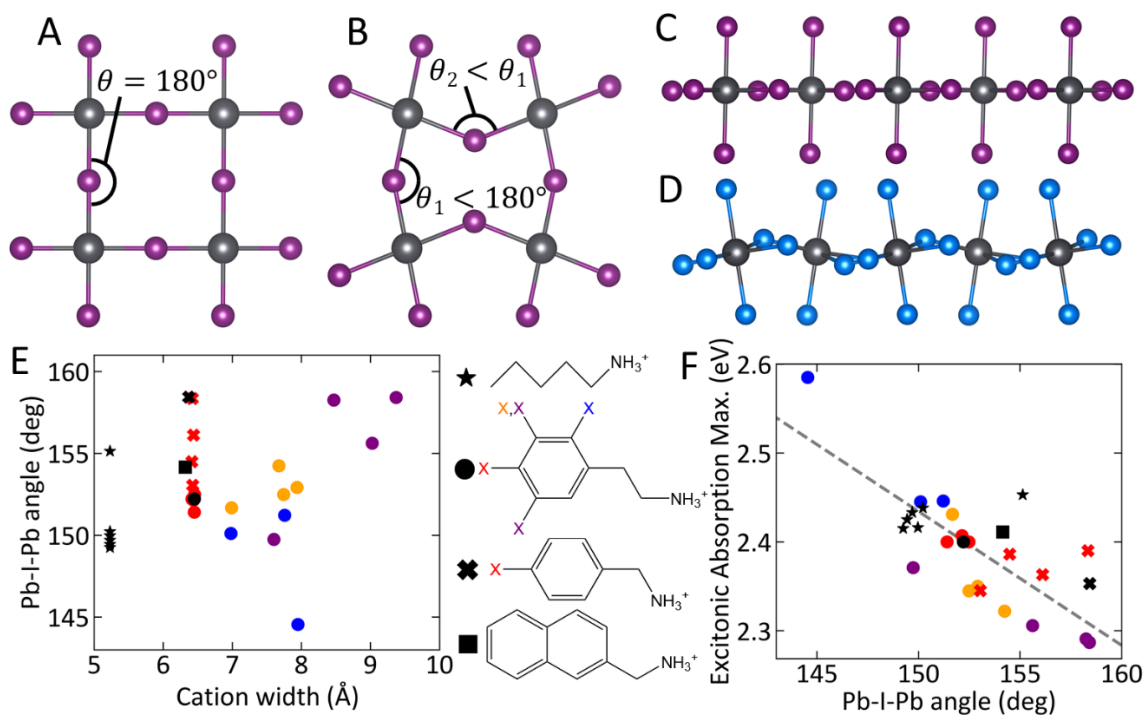

**Figure 4: Distortion in the inorganic framework.** A) Top-view of a perfect cubic 2DHP, with the Pb-I-Pb angle $\theta = 180°$. B) Top-view of a distorted 2DHP, where $\theta_2 < \theta_1 < 180°$. C) Planar and D) corrugated inorganic frameworks. E) Scatter plot of cation width and $\theta$ for experimental Pb-I 2DHPs, where the symbol shape corresponds to the cation family and color is used to specify the type of substitution. F) Scatter plot of room-temperature excitonic absorption maximum and Pb-I-Pb angle for experimental 2DHPs. Linear fit excluding $(C_nH_{2n+1}NH_3)_2PbI_4$ (stars) is shown in grey. Data in E and F from refs (26, 37, 59–64).

## 3.1 Cation-induced structural changes

A large library of cations can be used to synthesize 2DHPs since there is only a restriction on its cross-sectional area.(25) We highlight how the geometry, i.e., the length and width, and the composition, that drives chemical interactions within the organic and between the organic and inorganic layers, of the organic cation induce structural changes in the 2DHPs.(26, 37, 59, 60, 65) We examine several families of cations and discuss how modifications to the cation affect the inorganic framework and emphasize that this is not an exhaustive list of lead iodide-based 2DHPs. We point the reader to ref (25) for a comprehensive list of 2DHPs indexed in the Cambridge Structural Database. The width of the cation is quantified by measuring the distance in the reported crystal structure between atoms orthogonal to the alkylammonium tether and adding the appropriate van der Waals radii(66) of the terminal atoms. For example, we quantify the width of PEA cations and its derivatives as the distance between the two *ortho* or the two *meta* substituents on the phenyl group. If the Pb-I framework is asymmetric such that there are two or more distinct Pb-I-Pb bond angles $\theta$ ($\theta_2 \neq \theta_1$ in Figure 4B), we report the average $\theta$.

Alkylammonium ($C_nH_{2n+1}NH_3^+$) cations behave as one would expect—as the length of the cation gets longer (i.e., $n$ increases), there is no regular change to the degree of distortion present in the inorganic framework because the width of the cation remains unchanged (stars, Figure 4E). The structural similarity is seen for $n = 4,5,7,8,9$. Interestingly, $n = 6$ is the exception and is significantly less distorted with a ~5° greater Pb-I-Pb bond angle than cations with $n \neq 6$.(65) For $n \neq 6$, there is a sharp discontinuity in the temperature-dependent excitonic absorption spectra, where the excitonic resonance abruptly red-shifts upon warming between 235 and 310 K; this discontinuity is absent for $n=6$, even though there is still a structural phase transition at 258 K.(47, 63, 65) Despite the difference in their optical behavior, these are all first-order phase transitions.(63, 64)

2DHPs with derivatives of the PEA cation exhibit different behavior depending on the class of structural modification performed on the cation. Unmodified PEA is shown as the black circle in Figure 4E.(37) The length of the PEA cation can be increased without affecting the width by replacing the 4-position hydrogen (*para* to the ethylammonium tether) with a larger atom or functional group, such as a halogen or methyl group (red circles, Figure 4E). These modifications increase the spacing between inorganic layers but do not significantly alter the inorganic framework, which can be seen by their very similar Pb-I-Pb angles.(37)

Replacing the 2-position hydrogen (*ortho* to the ethylammonium tether) on the PEA cation with a larger substituent results in wider cations (blue circles, Figure 4E), and the inorganic frameworks of these 2DHPs are more distorted than that of unmodified $(PEA)_2PbI_4$.(59) The trend is not fully linear with size, however, as the 2-ClPEA cation is wider than the 2-FPEA cation yet results in a less distorted inorganic framework. The 2-Br cation results in the most distorted inorganic framework included in the plot with an average Pb-I-Pb angle of 144.55°, and unlike the other $(PEA)_2PbI_4$ derivatives that have planar inorganic frameworks (Figure 4C), its inorganic framework is corrugated (Figure 4D).

Wider cations do not always distort the inorganic framework, however. The orange circles in Figure 4E represent $(PEA)_2PbI_4$ derivatives where the 3-position hydrogen (*meta* to the ethylammonium tether) is replaced with a halogen or methyl group. While $(3-XPEA)_2PbI_4$ (X=F) is more distorted than unsubstituted $(PEA)_2PbI_4$,(62) the inorganic framework becomes progressively less distorted as the width of the cation is further increased (X=Cl, Br, Me), and these three 2DHPs are less distorted than unsubstituted $(PEA)_2PbI_4$.(60) Replacing both the 3- and 5- position *meta* hydrogens on the PEA cation with F, Cl, Br, or Me (purple circles, Figure 4E) results in a similar trend, with (3,5-FPEA)$_2$PbI$_4$ being more distorted than unsubstituted (PEA)$_2$PbI$_4$ and the others being less distorted.(60)

Replacing the ethylammonium tether on the PEA cation with a methylammonium tether results in the phenylmethylammonium (C$_6$H$_5$CH$_2$NH$_3^+$, abbreviated as PMA) cation, and (PMA)$_2$PbI$_4$ (black cross, Figure 4E) is significantly less distorted than (PEA)$_2$PbI$_4$.(61) Replacing the 4-position (*para*) hydrogen on the PMA cation with a larger atom or functional group (red crosses, Figure 4E) results in a very different trend than that observed with 4-XPEA cations. (4-FPMA)$_2$PbI$_4$ displays a similar degree of distortion to unsubstituted (PMA)$_2$PbI$_4$, but (4-XPMA)$_2$PbI$_4$ (X=Cl, Br, I) are significantly more distorted, likely because of halogen bonding interactions between the halogen on the cation and the axial I atoms in the inorganic framework.(26) Ruddlesden-Popper and Dion-Jacobson perovskites are both found in this family of 2DHPs (Figure 3D), and the stacking motif is tuned by halogen bonding interactions between cations in adjacent layers.(26)

We hypothesize that the differences in the degree of distortion between (4-XPEA)$_2$PbI$_4$ and (4-XPMA)$_2$PbI$_4$ 2DHPs originates from the methylammonium tether on the PMA cation being less flexible than the ethylammonium tether on the PEA cation, and the longer cations and/or halogen bonding interactions between the cations impose additional stress on the shorter tether. This hypothesis is supported by the existence of the 2DHP 2-naphthylethylammonium lead iodide where the phenyl group in the PEA cation is replaced by the longer naphthyl group (black square, Figure 4E).(67) Despite the size of the cation, 2-naphthylethylammonium lead iodide is less distorted than (PEA)$_2$PbI$_4$. However, the perovskite structure does not form if the 2-naphthylmethylammonium cation, which has a shorter alkylammonium tether, is used. Instead, an alternate structure-type forms that contains 1D chains of face-sharing lead iodide octahedra.(61)

2DHPs have also been synthesized with chiral cations.(68) Some chiral 2DHPs are optically active, where they preferentially absorb and emit either left- or right-hand circularly polarized light.(69–73)

**3.2 Effect of distortion on optical properties**

The degree of distortion in the inorganic framework of a 2DHP directly affects its optical properties.(51, 59, 67) We only consider the excitonic and continuum optical properties in this review and do not discuss the broad lower-energy photoluminescence resonance responsible for white light emission that is present in many 2DHPs. We suggest refs (74–77) for theoretical and experimental analysis of this emission feature.

As the inorganic framework of the 2DHP becomes more distorted, the band gap increases. Figure 4F plots the energy of the room-temperature excitonic absorption maximum against the Pb-I-Pb bond angle for the same 2DHPs included in Figure 4E. Because of the difficulty in determining the optical band gap at room temperature, we use the excitonic absorption maximum as a proxy. This is a valid approximation for cations with similar dielectric constants because the exciton binding energies will be similar (Sections 2.3 and 2.5). The grey line in Figure 4F shows the result of a linear regression performed on the 2DHPs with arylammonium cations, excluding those with alkylammonium cations (stars, Figures 4E-F). There is a strong correlation between the excitonic absorption maximum and the degree of distortion ($R^2 = 0.61$) that is highly statistically significant ($p = 1.6 \times 10^{-5} \ll 0.05$).

Knutson et al. analyzed the changes in orbital overlap for Sn-I 2DHPs, and the same arguments apply for Pb-I 2DHPs.(51) Increasing the Pb-I-Pb angle decreases the degree of orbital overlap between the metal *s* orbitals and halide *p* orbitals, lowering the energy of the top of the valence band. While a similar effect would also lower the top of the conduction band, the distortions break the symmetry of a perfect perovskite lattice and increase the antibonding interaction between metal *p* and halogen orbitals. This increases the energy of the conduction band, also contributing to the widening of the band gap.

### 3.3 Effect of temperature on distortions and optical properties

While increasing the degree of distortion in the inorganic lattice directly correlates with a widening band gap, a more complicated picture emerges when considering temperature-dependent changes. Figure 5 shows the temperature-dependent structural and optical properties of $(PEA)_2PbI_4$. Structural data is determined by single-crystal X-ray diffraction measurements. The 100 K and 300 K structures were previously reported in ref (37), and the 125-275 K structures are available at the Cambridge Crystallographic Data Centre under deposition codes 2097583-2097589.

A common feature of oxide and halide perovskites is that they become less distorted and the unit cell volume increases as the temperature increases,(12) and this is also observed for $(PEA)_2PbI_4$. The Pb-I-Pb bond angles increase and the inorganic framework becomes less distorted as the 2DHP is warmed (Figure 5A), which based on the discussion in Section 3.2 should lead to a reduction in band gap. In most semiconductors, the band gap decreases in energy as the temperature is increased. However, here the opposite temperature dependence is observed: the excitonic absorption maximum *increases* in energy upon warming (Figure 5B.(37, 52) This phenomenon is also seen in 3DHPs as well as in lead chalcogenides.(78, 79)

In the 3DHP methylammonium lead iodide, the blue-shift upon warming was thermodynamically attributed to the large increase in unit cell volume as the temperature rises.(79) We believe that this phenomenon warrants further study because the structural changes are more subtle than what can be inferred by solely examining the volume. While the volume of the inorganic framework increases, which can be seen by the increase in *a* and *b* lattice constants (Figure 5C), virtually all the volume increase can be attributed to an increase in Pb-I-Pb bond angle because the Pb-I bonds do not significantly elongate as the temperature increases (Figure 5D). The lack of Pb-I bond elongation is not unique to $(PEA)_2PbI_4$. Indeed, in the 3D inorganic halide perovskite $CsPbI_3$, the Pb-I bond lengths decrease as the temperature increases, even though the overall unit cell volume increases.(17, 80) We hypothesize that the widening band gap as the temperature increases is related to the unusual thermal expansion properties, and a careful theoretical analysis is needed to characterize the origins of these structural and electronic phenomena.

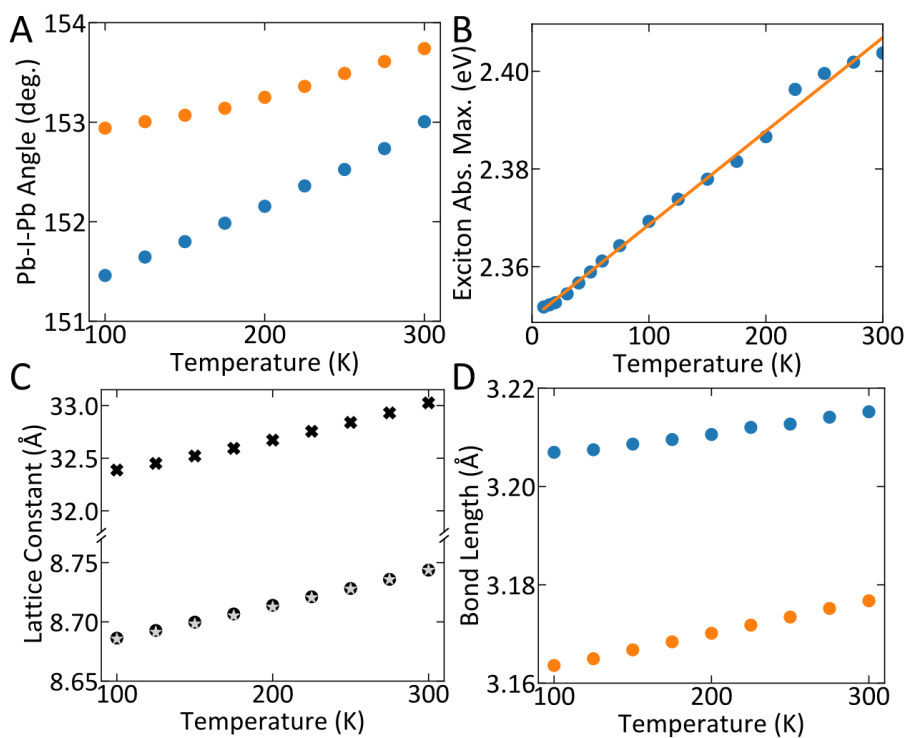

**Figure 5: Temperature-dependent structural properties of $(PEA)_2PbI_4$.** Temperature-dependent change in A) Pb-I-Pb bond angles $\theta_1$ (orange) and $\theta_2$ (blue); B) excitonic absorption maximum (blue) with linear regression (orange); C) *a* (circles), *b* (stars), and *c* (crosses) lattice constants; D) equatorial (orange) and axial (blue) Pb-I bond lengths.

## 4. Energetic disorder caused by the free volume in the organic framework

Even when the cation does not induce significant structural distortion leaving the exciton binding energy and band gap unchanged, significant changes in the optical spectra can be found. For example, in the (4-XPEA)$_2$PbI$_4$ (X=H, F, Cl, Br, CH$_3$) family (Figures 4E-F, black and red circles), lengthening the PEA cation by introducing a larger substituent in the 4-position of the phenyl group increases the interlayer Pb-Pb spacing (*d*, Figure 1C-D). However, it does not significantly distort the inorganic framework, as seen by the invariance in the Pb-I-Pb bond angle and excitonic absorption maximum (black and red circles, Figure 4E-F).(37)

As the cation is lengthened, the excitonic absorption resonance significantly broadens at both room temperature (Figure 6A) and at 10 K (Figure 6B).(37) The broadening implies that longer (but not heavier, since CH$_3$ is lighter than Cl) cations create significant energetic disorder. To quantify the effect of cation substitution on the excitonic absorption resonance, the temperature-dependent linewidth (Figure 6C) is fit to a model that accounts for vibration-induced linewidth broadening developed for transition-metal dichalcogenides(81) but is also applied in 2D and 3DHPs.(37, 82–84) The linewidth $\Gamma(T)$ is fit to the equation

$$\Gamma(T) = \Gamma(0) + \frac{\gamma_{LO}}{\exp\left(\frac{E_{LO}}{k_B T}\right) - 1} \qquad (\text{Eq. 2})$$

where $\Gamma(0)$ is the zero-temperature linewidth, $\gamma_{LO}$ is the electron-phonon coupling strength to LO phonons, and $E_{LO}$ is the energy of the phonon that couples to the exciton (Figure 6D). $\Gamma(0)$ increases as the cation length increases, raising the effective temperature of the 2DHP. In addition, $\gamma_{LO}$ significantly decreases as the cation gets longer, indicating that longer cations reduce the electron-phonon coupling strength. The phonon energy $E_{LO}$ will be discussed at length in Section 5.

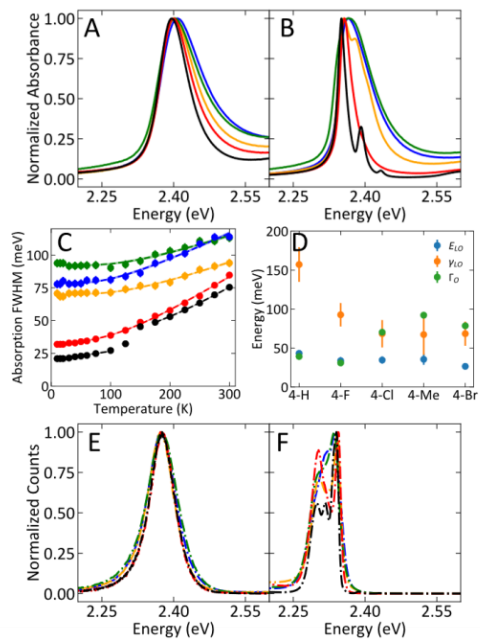

**Figure 6: Longer cations and energetic disorder**. Absorption spectra at A) 300 K and B) 10 K of (4-XPEA)$_2$PbI$_4$ (X=H (black), F (red), Cl (orange), Br (blue), CH$_3$ (green), with C) temperature-dependent full-width at half-maxima (FWHM) linewidths and D) extracted fit parameters. Photoluminescence spectra at E) 300 K and F) 10 K. Adapted with permission from ref (37). Copyright 2019 American Chemical Society.

Structurally, it was shown that longer cations increase the volume available to the cation as larger cations increase the interlayer spacing, but do not fill all the additional interstitial space on its own because the 4-XPEA cations are identical in width (black and red circles, Figure 4E). The additional free-volume in the organic layer reduces the degree to which the organic cations physically confine the inorganic layers.(37) The absorption process samples the instantaneous ground state energy landscape of the 2DHP and reflects the additional disorder induced by the longer cations. This hypothesis is further supported by the observation that a more rigid organic layer increases the photoluminescence quantum yield and reduces nonradiative recombination because additional motion of the organic and/or inorganic frameworks may create more nonradiative recombination pathways.(85)

In contrast to the excitonic absorption resonance, photoluminescence spectra show no significant increase in linewidth with changes in cation length (Figure 6E and F). Polaron formation (*i.e.*, electron-phonon coupling) in the excited state is hypothesized to stabilize a particular lattice configuration. Excitons in 2DHPs live for several picoseconds before undergoing photoluminescence, and accordingly polaron formation funnels the excitons to an energetic minimum before they recombine.(37) The difference between these absorption and photoluminescence spectra demonstrate that the ground and excited states of 2DHPs behave differently, which is an important fact to consider when modeling their structural, optical, and electronic properties.

Other cation families with tunable length do not show length-dependent changes to their optical spectra. For instance, the excitonic absorption spectra of alkylammonium lead iodide (($C_nH_{2n+1}NH_3$)$_2$PbI$_4$) 2DHPs (Figure 3A) do not broaden as the length of the cation changes.(47) We hypothesize that the difference in behavior between the ($C_nH_{2n+1}NH_3$)$_2$PbI$_4$ and (4-XPEA)$_2$PbI$_4$ 2DHP families originates from the difference in cation geometry, where unlike (4-XPEA)$_2$PbI$_4$, alkylammonium cations have a constant width over their entire length, and thus do not create additional free volume as their length increases.

## 5. Structure in Excitonic Optical Spectra and Hot Exciton Photoluminescence

At temperatures below 100 K, the excitonic absorption and photoluminescence resonances in the archetypal 2DHP (PEA)$_2$PbI$_4$ split into several discrete resonances.(46, 52, 86) This section investigates the origin of this structure as well as how the changes to the cation modify the spacing of the resonances.

### 5.1 Cryogenic optical spectra

Figure 7A shows the excitonic absorption (black) and photoluminescence (blue) spectra at a temperature of 15 K.(52) In the absorption spectrum, there are three discrete resonances: $Ex_{1a}$ at 2.355 eV, $Ex_{1b}$ at 2.398 eV, and $Ex_{1c}$ at 2.438 eV, that are separated from one another by 40-43 meV. These resonances fit to Lorentzian lineshapes, indicating that they are homogeneously broadened (*i.e.*, all excitons experience the same local environment).

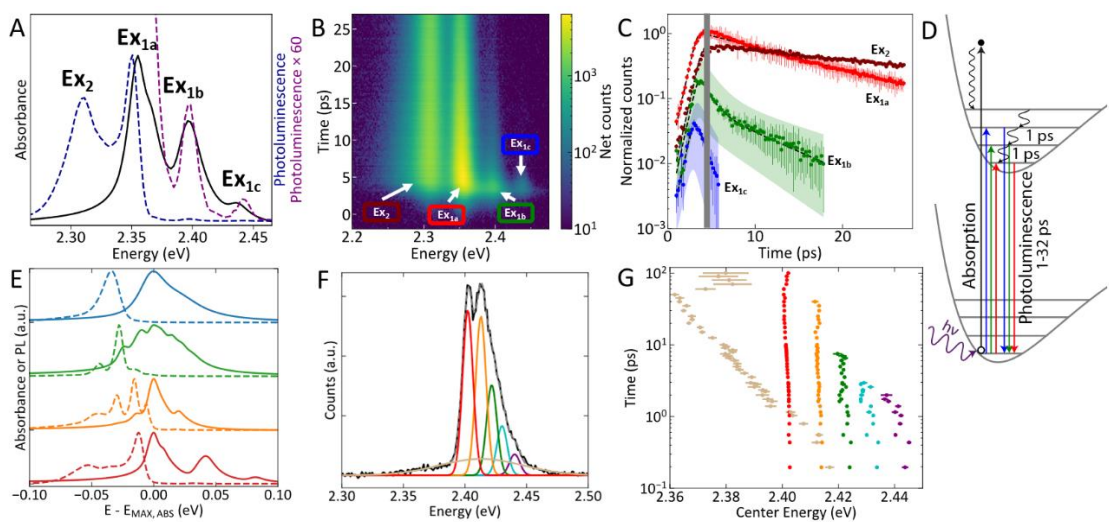

**Figure 7: Discrete excitonic structure in (PEA)$_2$PbI$_4$.** A) 15 K absorption (black) and photoluminescence (blue, ×60 in purple), and B) time-resolved photoluminescence. C) Lifetime fit of integrated amplitude of photoluminescence resonances shown in B). D) Schematic detailing low temperature absorption and photoluminescence processes for *Ex$_{1a}$* (red), *Ex$_{1b}$* (green), and *Ex$_{1c}$* (blue). Non-radiative relaxation between vibrational states is shown by black curved arrows. A-D adapted with permission from ref (52). Copyright 2016 American Chemical Society. E) 10 K absorption (solid) and photoluminescence (dashed) of (PEA)$_2$PbI$_4$ (red) and (2-XPEA)$_2$PbI$_4$ (X=F (orange), Cl (green), Br (blue). F) Fit of ultrafast time-resolved photoluminescence spectrum of (2-ClPEA)$_2$PbI$_4$ taken at *t* = 0.55 ps. G) Energies of resonances found by fitting each time step of the time-resolved photoluminescence spectrum of (2-ClPEA)$_2$PbI$_4$. E-G reprinted with permission from ref (59). Copyright 2020 American Chemical Society.

In photoluminescence, *Ex$_{1a}$* is present along with another strong Stokes-shifted resonance, *Ex$_2$*, which is separated from *Ex$_{1a}$* by 40 meV. When the photoluminescence spectrum is magnified, two additional resonances appear (purple, Figure 7A) at the same energies *Ex$_{1b}$* and *Ex$_{1c}$* that are seen in absorption.(52) These hot photoluminescence resonances are noteworthy because in most emitters carriers lose their excess kinetic energy through vibrational relaxation to the lowest-lying excited state before recombining; this principle is known as Kasha's rule.(87) 0.3% of the photoluminescence in (PEA)$_2$PbI$_4$ occurs out of these hot states.(59) In contrast to the excitonic absorption resonances, the resonances in photoluminescence are best fit with Gaussian functions, indicating that the resonances are inhomogeneously broadened.(52) Each exciton therefore occupies a different local environment when recombining.

In Sections 5.2-5.4, we describe hypotheses for the origin of excitonic absorption and photoluminescence resonances labeled *Ex$_{1a}$*, *Ex$_{1b}$*, and *Ex$_{1c}$*. The photoluminescence peak *Ex$_2$* is reported to be a bound exciton(85) or biexciton emission(88–90) as *Ex$_2$* shows a different intensity dependence than *Ex$_{1a}$*. *Ex$_{1a}$* scales linearly with excitation density(59) until saturation occurs, and the saturation behavior is hypothesized to be caused by exciton-exciton annihilation (analogous to Auger recombination for free carriers)(85) or biexciton formation.(88) In contrast, *Ex$_2$* saturates under continuous illumination,(59) while it grows linearly(85) or superlinearly(88) when excited with ultrafast pulsed excitation. We note that there is an additional shoulder in the absorption spectrum, which is consistent with a resonance 14 meV higher in energy, that we do not discuss here.(52)

### 5.2 The exciton-phonon coupling hypothesis

One hypothesis is that *Ex$_{1a}$*, *Ex$_{1b}$*, and *Ex$_{1c}$* are caused by the coupling of the exciton to a vibrational mode.(52, 59, 91, 92) Ultrafast time-resolved photoluminescence measurements (Figure 7B-C) are used to observe the fate of photoexcitations in 2DHPs, monitoring the population of the individual excitonic resonances. The population of $Ex_{1c}$ decays while that of $Ex_{1b}$ continues to rise (grey line, Figure 7B), and likewise the population of $Ex_{1b}$ begins to decay while that of $Ex_{1a}$ and $Ex_2$ continue to increase. The kinetics suggest that a selection rule operates such that excitons cannot relax directly into all states, but instead can only relax from the *n*th to the *n-1*th state, like in the quantum harmonic oscillator (Figure 7D). The equal spacing between resonances combined with the photoluminescence kinetics led us to hypothesize that $Ex_{1b}$ and $Ex_{1c}$ are vibronic replicas of $Ex_{1a}$, caused by the exciton coupling to a ~40 meV phonon.(52) Theoretical calculations show that a phonon with this energy must exclusively involve motion of the cation because modes involving the heavy Pb and I atoms are all much lower in energy, and the phonon likely involves a twisting of the phenyl group.(52)

In addition to the splitting of the absorption and photoluminescence spectra and the photoluminescence kinetics, other evidence supports the assignment of a vibronic progression. Magnetooptical measurements show an identical shift with the magnetic field for all resonances, indicating that the discrete resonances all have the same origin.(92) As the temperature increases, $Ex_{1a}$, $Ex_{1b}$, and $Ex_{1c}$ in absorption merge into a single resonance (black, Figure 6E-F). Using Eq. 2 and fitting the exciton absorption linewidth of $(PEA)_2PbI_4$ from 175-300 K, it was found that the optical phonon that broadens the exciton has an energy $E_{LO}$ of 43 meV, matching the energetic spacing between $Ex_{1a}$, $Ex_{1b}$, and $Ex_{1c}$ at 15 K.(37) In addition, $(PEA)_2PbI_4$ exhibits strong electron-phonon coupling with a coupling constant $\gamma_{LO}$ of 160 meV, four times greater than that of lead iodide 3DHPs.(82) Fitting the photoluminescence spectrum from 125-300 K to Eq. 2 gives similar results, yielding a phonon energy $E_{LO}$ of 45 meV and an electron-phonon coupling constant $\gamma_{LO}$ of 190 meV. We separately extracted the absorption linewidth of $Ex_{1a}$ from lower, 10-100 K temperature spectra, where it is resolved separately from the thermally broadened absorption envelope that contains $Ex_{1a}$, $Ex_{1b}$, and $Ex_{1c}$. Fitting $Ex_{1a}$ at lower temperature, we found $E_{LO}$ is 6.9 meV and $\gamma_{LO}$ is 8 meV, showing that $Ex_{1a}$ on its own weakly couples to a low-energy phonon distinct from the phonon involved in the vibronic progression.(37) We note that other groups have fit the linewidth of the entire photoluminescence spectrum from ~10-300 K in a single regression and found $E_{LO}$ = 17 or 18 meV and $\gamma_{LO}$ = 25 or 43 meV,(83, 91) values that are in between those we found when separately fitting the high- and low-temperature regimes. We hypothesize that fitting the entire spectrum averages the low- and high-temperature regimes, which we do not believe to be accurate given that the discrete resonances are only resolved at temperatures below 100 K in $(PEA)_2PbI_4$.

**5.3 Testing the exciton-phonon coupling hypothesis through cation substitution**

The phonon that couples to the exciton was hypothesized to reside on the organic cation. Thus, modifying the cation is expected to alter the energy of the phonon that couples to the exciton. The energy of a normal stretching vibration is inversely proportional to the mass of the vibrating atoms. In a twisting mode, the mass is replaced with the moment of inertia, which quantifies how far the mass is from the rotational axis. Accordingly, if mass is added close to the rotational axis, the moment of inertia and therefore the energy of a twisting vibration will hardly change. If mass is added far from the rotational axis, the energy of the vibration will be greatly decreased.

We hypothesized that the rotational axis passes through the 1,4-axis of the PEA cation, parallel to the ethylammonium tether and synthesized and spectroscopically measured (4-XPEA)$_2$PbI$_4$ derivatives, adding mass to the rotational axis, and (2-XPEA)$_2$PbI$_4$ derivatives, adding mass away from the rotational axis, to identify the nature of the vibrational mode that couples to the exciton. While the increased energetic disorder in (4-XPEA)$_2$PbI$_4$ washes out the excitonic structure consistent with weaker electron-phonon coupling (Section 4), using Eq. 2 to fit the excitonic absorption linewidths we found that $E_{LO}$ decreases from 43 meV for unsubstituted (PEA)$_2$PbI$_4$ by 19% to 34-36 meV for X=F, Cl, Me, and by 37% to 27 meV for X=Br.(37) The changes in the phonon energy are small considering that the mass of the 4-position substituent increases from 1 u (unsubstituted PEA) to 85 u (4-BrPEA), suggesting the vibrational mode that couples to the exciton is a twisting mode of the cation.

The photophysics of (2-XPEA)$_2$PbI$_4$ derivatives show much more extreme changes with single-atom substitution.(59) Multiple excitonic resonances are observed in their low-temperature absorption and photoluminescence spectra, which we used to directly quantify the energy of the phonon that couples to the exciton (Figure 7E). Ultrafast photoluminescence measurements, akin to those in Figure 7B-C, were collected to monitor the kinetics and identify excitonic resonances and their vibronic replicas. In (2-FPEA)$_2$PbI$_4$ (orange, Figure 7E), the photoluminescence kinetics suggest that there are two distinct manifolds of states, with the resonances in each manifold separated from one another by 20-24 meV.(59) We hypothesized the presence of two manifolds of states is caused by asymmetry in the inorganic lattice, where there are two distinct Pb-I-Pb bond angles that differ by ~3° from one another.

In (2-ClPEA)$_2$PbI$_4$ (green, Figure 7E), we found one manifold of states where the states are separated by 12-13 meV from one another, consistent with the symmetric inorganic framework of (2-ClPEA)$_2$PbI$_4$ only having a single Pb-I-Pb bond angle, in contrast with the asymmetric lattice of (2-FPEA)$_2$PbI$_4$. The smaller spacing between resonances in (2-ClPEA)$_2$PbI$_4$ compared to in (2-FPEA)$_2$PbI$_4$ is consistent with the heaver Cl atom resulting in a larger moment of inertia and therefore a smaller vibrational energy than the 2-FPEA cation. In time-resolved photoluminescence spectra, we were able to identify five discrete states, four of which are hot (Figure 7F-G). The fraction of hot exciton photoluminescence in (2-FPEA)$_2$PbI$_4$ and (2-ClPEA)$_2$PbI$_4$ is 9.8% and 8.4% respectively, more than an order of magnitude greater than in unsubstituted (PEA)$_2$PbI$_4$. No excitonic structure was observed in (2-BrPEA)$_2$PbI$_4$ (blue, Figure 7E), which we hypothesize is either caused by the heavy Br atom reducing the phonon energy such that discrete resonances blend together or the highly-distorted, corrugated inorganic framework altering the optical properties. The much smaller splitting between resonances in (2-XPEA)$_2$PbI$_4$ supports the hypothesis that the excitonic structure is caused by coupling to a twisting mode of the PEA cation. Hot photoluminescence resonances separated from the central resonance by 15 meV were also found in (3-FPEA)$_2$PbI$_4$.(91)

**5.4 Criticism of electron-phonon coupling hypothesis**

The hypothesis that the excitonic structure in (PEA)$_2$PbI$_4$ and its derivatives is caused by coupling to a phonon is controversial(93) for several reasons. The first reason is that there is no mirror symmetry between the excitonic absorption and photoluminescence spectra. In a typical molecular vibronic progression, the absorption resonances are mirrored in photoluminescence about the zero-phonon line ($Ex_{1a}$ in Figure 7A).(94) This is mirror symmetry is imposed by the Franck-Condon approximation, which states that electronic transition probabilities do not depend on the nuclear coordinates.(95) According to the Franck-Condon principle, in (PEA)$_2$PbI$_4$ we would expect to observe replicas of $Ex_{1b}$ and $Ex_{1c}$ at lower in energy than $Ex_{1a}$ in photoluminescence, which is not what is observed. However, 3D lead halide perovskites are known to undergo structural changes to their lattice on a picosecond timescale upon photoexcitation.(96, 97) To the best of our knowledge, light-induced structural changes have not been studied in 2DHPs, but given the similarity to 3DHPs we expect this phenomenon to be present. Structural changes to the lattice would invalidate the Condon approximation, so mirror symmetry would no longer be expected.

The second major criticism of the vibronic coupling hypothesis is that in (PEA)$_2$PbI$_4$, $Ex_{1a}$ and $Ex_{1b}$ were found to couple differently to low energy (1-8 meV) phonons located on the inorganic framework. Figure 8 shows resonant impulsive stimulated Raman spectra when either $Ex_{1b}$ (Figure 8A) or $Ex_{1a}$ (Figure 8B) is excited.(93, 98) The 2D Raman spectra are generated by Fourier transforming the transient absorption spectra. If $Ex_{1b}$ were a vibronic replica of $Ex_{1a}$, it would be expected that the resonances would couple similarly to the inorganic lattice, yet the Raman spectra differ.(93) Instead, it was hypothesized that $Ex_{1a}$, $Ex_{1b}$, $Ex_{1c}$, and $Ex_2$ are separate but correlated excitons that each have different binding energies.(93, 98)

We suggest that the difference in coupling between $Ex_{1b}$ and $Ex_{1a}$ to the inorganic lattice can be explained by the different lifetimes of the individual resonances. At 15 K, photoluminescence from $Ex_{1b}$ (green, Figure 7C) has a lifetime of 2 ps, whereas photoluminescence from $Ex_{1a}$ has a lifetime of 13 ps.(52) As structural reorganization of the inorganic lattice likely occurs on a picosecond timescale, $Ex_{1b}$ likely decays before the lattice reorganization is complete. Therefore, $Ex_{1b}$ perceives a different inorganic lattice configuration than $Ex_{1a}$, and the vibrational spectrum as well as the exciton-phonon coupling dynamics may change as the lattice reconfigures.

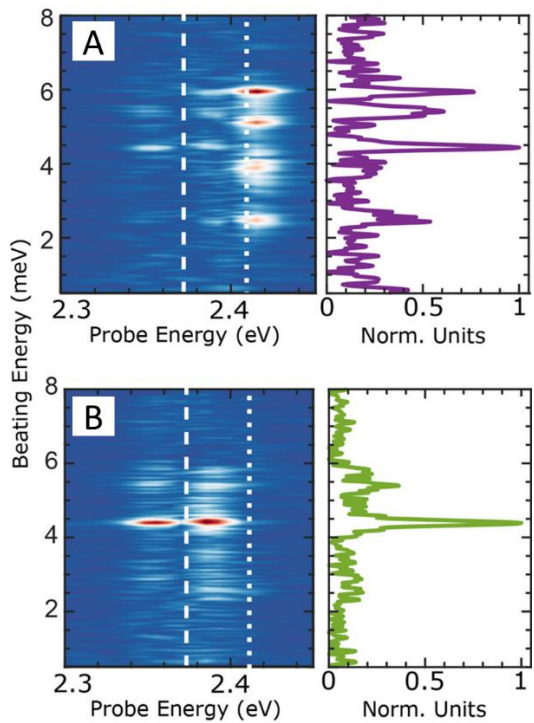

**Figure 8: Exciton-phonon coupling details.** 2D Resonant impulsive stimulated Raman scattering spectra (left) and integrated over detection energy (right) when A) $Ex_{1b}$ or B) $Ex_{1a}$ is excited. Reprinted with permission from ref (93). Copyright 2020 American Chemical Society.

We agree that $Ex_2$ likely has a different origin than $Ex_{1a}$, $Ex_{1b}$, $Ex_{1c}$ because of its different pump intensity dependence. However, it is too coincidental for $Ex_{1a}$, $Ex_{1b}$, and $Ex_{1c}$ to be distinct excitons with their own binding energy, yet they are separated from one another by a constant energy and their dynamics also obey the vibrational selection rule. Furthermore, the agreement of $E_{LO}$ from linewidth analysis with the spacing between $Ex_{1a}$, $Ex_{1b}$, $Ex_{1c}$ is consistent with the exciton coupling to a vibration on the organic cation. Finally, the ability to tune the excitonic structure through cation substitutions as well as the existence of five discrete excitonic resonances with equal separation in (2-ClPEA)$_2$PbI$_4$ (Figure 7F-

G) further supports the existence of a vibronic progression. 2DHPs have very complicated exciton dynamics, and further study is needed to confirm the origins of the low temperature excitonic structure.

## 6. Conclusion

A large library of organic cations can be introduced to synthesize stoichiometric crystal(lite)s of 2DHPs with tremendous diversity in structural,(25) optical, and electronic(34, 99) properties. The geometry (i.e., length and width) and the chemistry of the organic cation can alter the structure and dielectric environment of the inorganic framework, with most cations establishing a large spacing such that the inorganic layers are strongly quantum confined and yet weakly electronically coupled. 2DHPs have some unconventional optical properties in comparison to longer studied inorganic or organic semiconductors. Through systematic modifications of the organic cation, the community has synthesized, structurally characterized, and spectroscopically studied semiconducting 2DHPs to begin to uncover their unusual, yet fascinating photophysics. Even for aliphatic or short aromatic ammonium cations that create Type I heterostructures in which excitons reside in the inorganic framework, "at a distance" the organic cation can modify the energy, linewidth, and population dynamics of excitonic states. Near the band gap energy, low-temperature optical spectra of many 2DHPs show multiple excitonic resonances with ultrafast, picosecond photoluminescence lifetimes that are comparable to vibrational relaxation timescales, consistent with a breakdown in the Franck-Condon approximation, and give rise to hot exciton luminescence, in violation of Kasha's rule.

As scientists, we make wish lists about capabilities and knowledge. In 2DHPs, it would be nice to carry out single crystal diffraction measurements at the ~10-15 K temperatures where thermal broadening and thus linewidths in optical spectra are reduced to resolve multiple excitonic resonances. This would allow more detailed correlation of structure and properties and the identification of 1$^{st}$ and 2$^{nd}$ order phase transitions, and it would build confidence in spectral assignments. Advances in optical and vibrational spectroscopies have allowed greater sophistication in our ability to probe materials(95) and are being used to study 2DHPs. Uniting the community and its capabilities to map the similarities/differences in the energy, intensities, and timescales across techniques promises to address the differences in our observations and controversies in our analyses about excitonic and vibrational processes. Collaboration between experiment and theory to measure and model 2DHPs, particularly the dynamics in the excited state, is important in these materials that are soft and known to reorganize in response to carriers and excitons.

The relationships between structure and properties is important in establishing design rules that allow us to exploit the chemical diversity of these 2DHPs and their large exciton binding energy, which make them excellent candidates for excitonic devices. These devices include heterojunction solar cells and photodetectors,(100, 101) efficient fluorophores and light emitting diodes (LEDs) that emit spectrally narrow(102, 103) or broad-spectrum white(104) light, as the gain medium in lasers,(7, 105) and narrow bandwidth optical modulators. Despite the large binding energy, the photoluminescence quantum yield rapidly decays at temperatures above 200 K in many 2DHPs.(46, 106) In Type I 2DHPs,(34, 107) efficient electroluminescence has only been achieved at temperatures <200 K.(102, 103) The decrease in radiative recombination near room temperature has been attributed to thermal dissociation of excitons and strong electron-phonon coupling leading to rapid nonradiative recombination.(3, 46, 102, 103, 106, 108) Fully understanding the relationship between the structure of 2D perovskites and the resulting electronic/excitonic coupling to phonons is important to realize their promise as emissive materials. Slow carrier cooling in 2DHPs may allow hot carriers to be extracted before relaxing to the lowest-lying state(109, 110) and, it may be possible to further slow excitonic relaxation by tuning the cation. Slowing vibrational relaxation may also allow for population inversion and make 2DHPs suitable gain material in lasers.(82, 105) The atomic-scale perfection of the inorganic framework and the large binding energy in absorption suggests 2DHPs should enable ultra-sharp optical modulators,(111) but its realization requires minimizing the sources of energetic disorder that broaden the absorption linewidth. In addition to conventional devices, these unique properties make 2D perovskites a promising platform for exotic devices that use excitonics for energy transfer and communication such as a photonic wire that harnesses the energetic properties of excitons for directional transmission of energy(112) or novel polaritonic devices,(113) such as optical transistors and circuits that require strong exciton-photon coupling.(114)

**Acknowledgments**

This work is supported by CRK's Stephen J. Angello Professorship.